# Demonstrating Invalidity of Solution Methods Involved in Solving Compatibility Equations

Peng SHI[1, 2] *

[1] Logging Technology Research Institute, China National Logging Corporation, No.50 Zhangba five road, Xi'an, 710077, P. R. China

[2] Well Logging Technology Pilot Test Center, China National Logging Corporation, No.50 Zhangba five road, Xi'an, 710077, P. R. China

*Corresponding author: sp198911@outlook.com;

***Abstract.*** In the theory of elasticity, the constraint of compatibility conditions on displacement field ( $\nabla \times (\nabla \boldsymbol{u} + \boldsymbol{u}\nabla) \times \nabla = 0$ ) is not equivalent to the property of displacement field ( $\nabla \times (\nabla \boldsymbol{u}) = 0$ ). The difference may broaden the possibility of solutions to elasticity problems, which means that the solution methods involved in solving the compatibility equations like stress-based solution method and stress function method may be invalid. The study argues the validity of the solution methods involved in solving the compatibility equations. It is concluded that the absence of constraints on local rigid body rotation causes the non-uniqueness of displacement fields for a rectangular beam purely bent and a cantilever loaded at the end.

# 1. Introduction

The theory of elasticity studies the stress state and deformation of elastomer under various loads. The theory is continuously developed and improved in the process of solving practical engineering problems. Galileo first looked into the bending problem of beams in 1638, driven by the demands of building projects [1, 2]. In 1678, Hooke published the physical law, now known as Hooke's law. It states that the deformation of an elastic body is proportional to the applied force based on experimental results of metal wires, springs, and cantilever beams [1, 2]. The general equation of the theory of elasticity was derived by Navier and Cauchy in 1821 and 1822, providing the theoretical framework for elasticity.

In the theory of elasticity, elastomers are considered to be perfectly continuous and are paid no attention to their molecular structure, which is called continuum hypothesis [3, 4]. The motion of material elements constituting an elastomer is believed to be the motion of particles, which can be described by Newton's three laws of motion or other mechanical principles related to and equivalent to them [1-7]. Continuum hypothesis allows for the description of internal force acting on every given surface element in the form of a field and the use of powerful methods of calculus to describe the equilibrium of a free body with an infinitesimal volume in an elastomer [3, 4]. In order to conveniently describe the force on the bounding surface of a free body and the equilibrium of the free body whose volume goes to zero under resultant force, the stress tensor is introduced into the theory of elasticity, which is a second-order symmetric tensor [5, 6]. Since stress are caused by strain in an elastomer, the strain tensor is introduced to describe the constitutive relation of an elastomer and the deformation at a certain point.

At present, scholars studying the theory of elasticity have proposed various analytical methods for solving simple elastic problems, like displacement solution method, stress-based solution method, and stress function method, etc. Because the number of components of stress tensor and strain tensor are more than the number of displacement components, the compatibility conditions are needed when the elastic problems are solved with the methods other than displacement solution method. Here, we divide the analytical methods in the theory of elasticity into displacement solution method and the solution methods involved in solving compatibility equations. It is seen that the solutions involved in solving compatibility equations, like a rectangular beam purely bent and a cantilever loaded at the end, cannot guarantee the uniqueness of the displacement field. The uniqueness of the displacement field is considered to be determined by the displacement constraints of beams related to its rigid body motion [1, 2]. However, it seems not the case.

The study aims to argue the validity of the solution methods involved in solving compatibility equations. The well-known solutions of a rectangular beam purely bent and a cantilever loaded at the end are analyzed. It is found that the displacement fields for a rectangular beam purely bent and a cantilever loaded at the end are non-unique and the absence of constraints on local rigid body rotation causes the non-uniqueness of displacement fields.

## 2. Stress-based solution method for planar problems

In plane coordinates (*xoy*), the basic equations for plane stress problems are as follows [1]:

(1) Equilibrium equation:

$$\begin{cases} \dfrac{\partial \sigma_{xx}}{\partial x} + \dfrac{\partial \sigma_{xy}}{\partial y} + \dfrac{\partial V}{\partial x} = 0 \\ \dfrac{\partial \sigma_{xy}}{\partial x} + \dfrac{\partial \sigma_{yy}}{\partial y} + \dfrac{\partial V}{\partial y} = 0 \end{cases} \quad (1)$$

where, $\sigma_{xx}$ and $\sigma_{yy}$ are the normal stress components on the *x* and *y* planes, $\sigma_{xy}$ is the shear stress component, and *V* is the body force potential.

(2) Geometrical equation:

$$\begin{cases} \varepsilon_{xx} = \dfrac{\partial u}{\partial x} \\ \varepsilon_{yy} = \dfrac{\partial w}{\partial y} \\ \gamma_{xy} = \dfrac{\partial u}{\partial y} + \dfrac{\partial w}{\partial x} \end{cases} \quad (2)$$

where, $\varepsilon_{xx}$ and $\varepsilon_{yy}$ are the normal strain components on the *x* and *y* planes, $\gamma_{xy}$ is the shear stress component, *u* and *w* are the components of displacement along the *x* and *y* directions.

(3) Constitutive relationship:

$$\begin{cases} \varepsilon_{xx} = \dfrac{1}{E}\left(\sigma_{xx} - v\sigma_{yy}\right) \\ \varepsilon_{yy} = \dfrac{1}{E}\left(\sigma_{yy} - v\sigma_{xx}\right) \\ \gamma_{xy} = \dfrac{1}{G}\sigma_{xy} \end{cases} \quad (3)$$

where, *E* is the Young's modulus, *G* is the shear modulus and *v* is the Poisson's ratio.

(4) Compatibility equation:

$$\frac{\partial^2 \varepsilon_{xx}}{\partial y^2} + \frac{\partial^2 \varepsilon_{yy}}{\partial x^2} - \frac{\partial^2 \gamma_{xy}}{\partial x \partial y} = 0 \tag{4}$$

Submitting constitutive relationships into the compatibility equation of planar problems (Equation (4)) and utilizing the equilibrium equation (Equation (1)), the compatibility equation expressed with stress, which is known as B-M equation for planar problems, is obtained. In the case of plane stress, the B-M equation is expressed as [1, 2]:

$$\nabla^2 (\sigma_{xx} + \sigma_{yy}) = (1+v) \nabla^2 V \tag{5}$$

here, $\nabla^2$ is the two-dimensional Laplace operator. In the case of plane strain, the B-M equation is expressed as [1, 2]:

$$\nabla^2 (\sigma_{xx} + \sigma_{yy}) = \frac{1}{(1-v)} \nabla^2 V \tag{6}$$

In order to solving the equation, a function is often introduced to have the components of the stress tensor are represented as [1]:

$$\begin{cases} \sigma_{xx} = \frac{\partial^2 \phi}{\partial y^2} + V \\ \sigma_{yy} = \frac{\partial^2 \phi}{\partial x^2} + V \\ \sigma_{xy} = -\frac{\partial^2 \phi}{\partial x \partial y} \end{cases} \tag{7}$$

here, the function $\phi$ is called Airy stress function. Submitting Equation (7) into Equations (5) and (6), the basic equations of stress function method for plane stress and plane strain problems are obtained, respectively:

$$\nabla^2 \nabla^2 \phi = -(1-v) \nabla^2 V \tag{8}$$

$$\nabla^2\nabla^2\phi = -\frac{1-2v}{1-v}\nabla^2 V \tag{9}$$

When body forces in Equations (8) and (9) are constant, Equations (8) and (9) both simplified as:

$$\nabla^2\nabla^2\phi = 0 \tag{10}$$

Equation (10) are applied to some examples of practical interest.

### 3. Solutions to bending of rectangular beam under different loads

The bending of rectangular beam under different loads has been deeply studied in materials mechanics and elastic theory. In this section, we verify the validity of the solutions involved in solving compatibility equation using the pure bending of a rectangular beam and the bending of a cantilever loaded at the end as examples.

### 3.1 Solution to pure bending of rectangular beam

Considering a rectangular beam having a narrow cross section of unit width bent under a constant bending moment (Figure 1). The upper and lower edges are free from loads. Assuming the stress function for pure bending of rectangular beam is as follows [1]:

$$\phi = \frac{M}{4c^3} y^3 \tag{11}$$

with $M$ the constant bending moment and $c$ the half height of rectangular beam, the stress components are obtained:

$$\begin{cases} \sigma_{xx} = \frac{3M}{2c^3} y \\ \sigma_{yy} = 0 \\ \sigma_{xy} = 0 \end{cases} \tag{12}$$

Submitting Equation (12) into Equation (3), the strain components are obtained:

$$\begin{cases} \varepsilon_{xx} = \dfrac{M}{EI} y \\ \varepsilon_{yy} = -\dfrac{vM}{EI} y \\ \gamma_{xy} = 0 \end{cases} \quad (13)$$

with $I=2c^3/3$. Then, with Equation (2) and Equation (13), the displacement components are obtained [1]:

$$\begin{cases} u = \dfrac{M}{EI} xy - \omega y + u_0 \\ w = -\dfrac{vM}{2EI} y^2 - \dfrac{M}{2EI} x^2 + \omega x + w_0 \end{cases} \quad (14)$$

where, $\omega$, $u_0$ and $w_0$ are the integration constants.

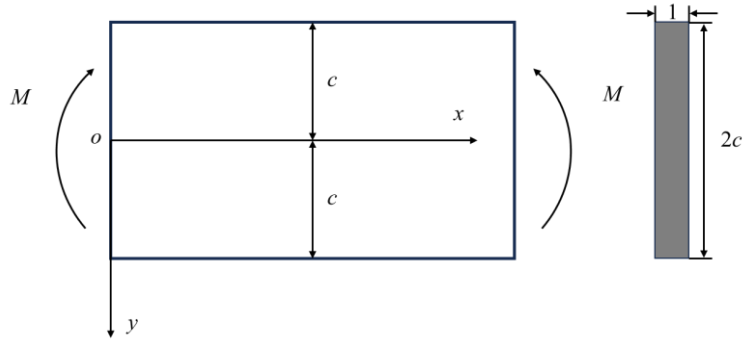

**Figure 1. Sketch of the pure bending of rectangular beam**

Equations (12)-(14) are the solution to pure bending of rectangular beam. It is seen that there are three undetermined constants ($\omega$, $u_0$ and $w_0$) in Equation (14). The three undetermined constants are considered to be determined by the displacement constraints of beams related to its rigid body motion [1]. $u_0$ and $w_0$ in Equation (14) are constant terms, therefore, $u_0$ and $w_0$ should represent the rigid body translation of rectangular beam. According to Equation (14), the displacement component $u$ is zero at the location of neutral layer ($y=0$), which is equal to the displacement component $u$ of beam's neutral layer when the beam does not rotate. This means that $\omega$ should be independent of rigid body motion. The term including $\omega$ in the displacement is independent of deformation, therefore the term should be related to local rigid body rotation. Since the local rigid body rotation is not taken into account for the compatibility conditions, the portion of displacement caused by local rigid body rotation should be arbitrary, which means that the solution of displacement conflicts with the uniqueness of solution of elasticity. This indicates that the validity of the solution methods involved in solving compatibility equation is questionable.

**3.2 Solution to bending of a cantilever loaded at the end**

Let a beam of narrow rectangular cross section of unit width be bent by a force $P$ applied at the end (Figure 2). The upper and lower edges are free from loads, having a resultant $P$, are distributed along the end $x=0$. Supposing the stress components are as follows [8]:

$$\begin{cases} \sigma_{xx} = -\dfrac{Pxy}{I} \\ \sigma_{yy} = 0 \\ \sigma_{xy} = -\dfrac{P}{2I}\left(c^2 - y^2\right) \end{cases} \quad (15)$$

the strain components are obtained by submitting Equation (15) into Equation (3):

$$\begin{cases} \varepsilon_{xx} = -\dfrac{Pxy}{EI} \\ \varepsilon_{yy} = \dfrac{vPxy}{EI} \\ \varepsilon_{xy} = -\dfrac{P}{2IG}\left(c^2 - y^2\right) \end{cases} \quad (16)$$

with Equation (2) and Equation (16), the displacement is obtained [2]:

$$\begin{cases} u = -\dfrac{Px^2 y}{2EI} - \dfrac{vPy^3}{6EI} + \dfrac{Py^3}{6GI} + ey + g \\ w = \dfrac{vPxy^2}{2EI} + \dfrac{Px^3}{6EI} + dx + h \end{cases} \quad (17)$$

where, *e*, *g*, *d* and *h* are the undetermined constants. In the undetermined constants, *e* and *d* satisfy the following relationship:

$$e + d = -\dfrac{Pc^2}{2GI} \quad (18)$$

The four undetermined constants are also considered to be determined by the displacement constraints of beams related to its rigid body motion and Equation (18) [8].

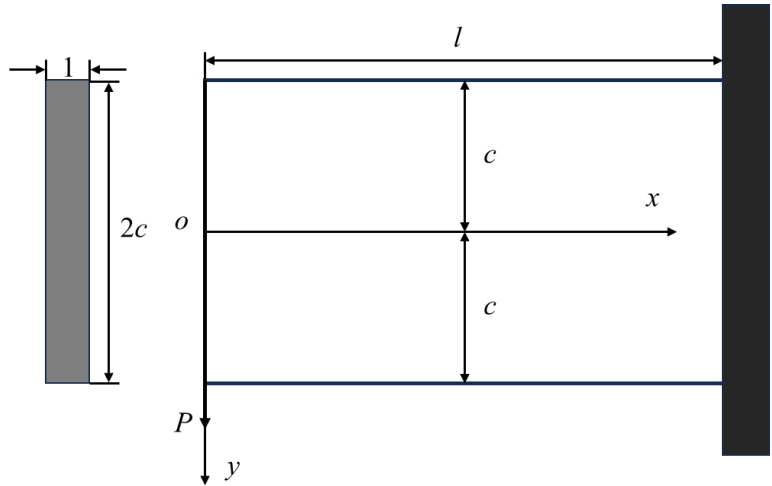

**Figure 2. Sketch of the bending of a cantilever loaded at the end**

It is not difficult to find from Equation (17) that *g* and *h* represent the rigid body translation and that the displacement component *u* is zero at the location of neutral layer (*y*=0), which is equal to the displacement component of beam's neutral layer when the beam does not rotate. This means that *e* and *d* should be independent of rigid body motion and be related to local rigid body rotation, which also supports that the solution methods involved in solving compatibility equations may be invalid.

## 4. Discussion

Submitting geometrical equations into compatibility equations, the compatibility equations expressed with displacement field are obtained:

$$\nabla \times (\nabla \boldsymbol{u} + \boldsymbol{u}\nabla) \times \nabla = 0 \qquad (19)$$

with $\boldsymbol{u}$ the displacement vector. According to Equation (19), the displacement field is a vector with a third derivative. However, it is not difficult to prove that Equation (19) changes the inherent properties of vector fields. Rewriting Equation (19), the following equation is obtained:

$$\nabla \times (\nabla \boldsymbol{u}) \times \nabla = -\nabla \times (\boldsymbol{u}\nabla) \times \nabla \qquad (20)$$

It is obtained from Equations (19) and (20) that $\nabla \times (\nabla \boldsymbol{u})$ and $(\boldsymbol{u}\nabla) \times \nabla$ described with stress or stress function may be non-zero. In this case, the solution methods involved in solving the compatibility equations should be invalid.

In classical theory of elasticity, although local rigid body rotation is acknowledged, the deformation is only taken into account for the compatibility conditions. In order to ensure the compatibility conditions of elastomer, the symmetric and antisymmetric parts of the gradient of displacement should be both considered. Since the continuous partial derivatives of multivariate functions are independent of the order of differentiation, the displacement field satisfies the following equation [9]:

$$\nabla \times (\nabla \boldsymbol{u}) = 0 \qquad (21)$$

which indicates that Equation (21) holds for an arbitrary vector with a second derivative.

Therefore, Equation (21) should be the compatibility conditions followed by the theory of elasticity. When the compatibility conditions do not consider the constraint of local rigid body rotation on the displacement field, the displacement part related to local rigid body rotation should be arbitrary. The study believes that the lack of constraints on local rigid body rotation is the root cause of non-uniqueness of displacement fields for a rectangular beam purely bent and a cantilever loaded at the end.

Assuming that a plane stress problem has been solved whose stress components are expressed with a stress function $\phi(x, y)$, the displacement components can be expressed with the stress function as follows:

$$\begin{cases} u = \dfrac{1}{E}\left(\int \dfrac{\partial^2 \phi}{\partial y^2} dx - v \dfrac{\partial \phi}{\partial x} + f_1(y)\right) \\ w = \dfrac{1}{E}\left(\int \dfrac{\partial^2 \phi}{\partial x^2} dy - v \dfrac{\partial \phi}{\partial y} + f_2(x)\right) \end{cases} \quad (22)$$

where, $f_1(y)$ and $f_2(x)$ are two undetermined functions. With Equation (22) and Equation (2), the following relationship is obtained:

$$\dfrac{G}{E}\left(\int \dfrac{\partial^3 \phi}{\partial y^3} dx + \int \dfrac{\partial^3 \phi}{\partial x^3} dy - 2v \dfrac{\partial^2 \phi}{\partial x \partial y} + \dfrac{\partial f_1(y)}{\partial y} + \dfrac{\partial f_2(x)}{\partial x}\right) = -\dfrac{\partial^2 \phi}{\partial x \partial y} \quad (23)$$

Rewriting undetermined functions $f_1(y)$ and $f_2(x)$ as follows:

$$\begin{cases} f_1(y) = g_1(y) + ay \\ f_2(x) = g_2(x) - ax \end{cases} \quad (24)$$

with $a$ an arbitrary constant, it is seen that no matter what value $a$ takes, Equation (23) always holds. This demonstrates that the lack of constraints on local rigid body rotation causes the non-uniqueness of displacement fields when the problems of elastic deformation are solved using methods involved in solving the compatibility equations.

## 5. Conclusions

The study demonstrated the invalidity of the methods involved in solving compatibility equations. It is found that the displacement fields are non-unique for the well-known solutions to pure bending of a rectangular beam and bending of a cantilever loaded at the end, and the portions of displacement generated by local rigid body rotation are arbitrary. The study comes to the conclusion that when the problems of elastic deformation are solved using methods involved in solving the compatibility equations, the primary reason of the non-uniqueness of displacement fields is the absence of constraints on local rigid body rotation.

**Acknowledgments**

This work was supported by the scientific research and technology development project of China National Petroleum Corporation (2020B-3713).

**Declaration of competing interest**

The author declares that he has no known competing financial interests or personal relationships that could have appeared to influence the work reported in this paper.

**References**

[1] Lu, M., Luo, X.: foundations of elasticity. Tsinghua university press, Beijing (2001).

[2] Timoshenko, S.: History of strength of materials: with a brief account of the history of theory of elasticity and theory of structures. Dove Publications, New York (1983).

[3] Lai, W.M., Rubin, D.H., Krempl, E.: Introduction to continuum mechanics. Butterworth-Heinemann, Burlington (2009).

[4] Malvern, L.E.: Introduction to the Mechanics of a Continuous Medium. Prentice Hall, New Jersey (1969).

[5] Graff, K.F.: Wave Motion in Elastic Solids. Dover publications, New York (1975).

[6] Achenbach, J.D.: Wave propagation in elastic solids. Elsevier, New York (1973).

[7] Landau, L.D., Lifshitz, E.M., Kosevich, A.M., Pitaevskii, L.P.: Theory of elasticity: volume 7. Elsevier, Oxford (1986).

[8] Timoshenko, S., Goodier, J. N.: Theory of elasticity. McGraw-Hill, New York (1951).

[9] Qiu, Z.: A simple theory of asymmetric linear elasticity. World J. Mech. **10** (2020) 166.